\documentclass[epj]{webofc}
\usepackage[utf8]{inputenc}
\usepackage[varg]{txfonts}   
\usepackage{booktabs}
\usepackage{xcolor}
\definecolor{darkred}{rgb}{0.4,0.0,0.0}
\definecolor{darkgreen}{rgb}{0.0,0.4,0.0}
\definecolor{darkblue}{rgb}{0.0,0.0,0.4}
\usepackage[bookmarks,linktocpage,colorlinks,
    linkcolor = darkred,
    urlcolor  = darkblue,
    citecolor = darkgreen]{hyperref}
\usepackage{subfigure}
\wocname{EPJ Web of Conferences}
\woctitle{Lattice2017}
\newcommand{\MSb}{{\overline{\rm MS}}}

\begin{document}
%
\selectlanguage{english}
\title{%
Progress in computing parton distribution functions from the quasi-PDF approach
}
\author{%
\firstname{Constantia} \lastname{Alexandrou}\inst{1,2} \and
\firstname{Krzysztof} \lastname{Cichy}\inst{3,4}\fnsep\thanks{Speaker, \email{kcichy@amu.edu.pl}} \and
\firstname{Martha}  \lastname{Constantinou}\inst{5}\and
\firstname{Kyriakos} \lastname{Hadjiyiannakou}\inst{2}\and
\firstname{Karl} \lastname{Jansen}\inst{6}\and
\firstname{Haralambos} \lastname{Panagopoulos}\inst{1}\and
\firstname{Aurora} \lastname{Scapellato}\inst{1,7}\and
\firstname{Fernanda} \lastname{Steffens}\inst{6}
}

\institute{%
Department of  Physics,  University  of Cyprus,  POB  20537,  1678  Nicosia,  Cyprus
\and
The Cyprus Institute, 20 Kavafi Str., Nicosia 2121, Cyprus
\and
Goethe-Universit\"at Frankfurt am Main, Institut f\"ur Theoretische
Physik, Max-von-Laue-Strasse 1, 60438 Frankfurt am Main, Germany
\and
Faculty of Physics, Adam Mickiewicz University, Umultowska 85, 61-614 Pozna\'{n}, Poland
\and
Department of Physics,  Temple University,  Philadelphia,  PA 19122 - 1801,  USA
\and
John von Neumann Institute for Computing (NIC), DESY, Platanenallee 6, 15738 Zeuthen, Germany
\and
University of Wuppertal, Gaußstr. 20, 42119 Wuppertal, Germany
}
\abstract{%
  We discuss the current developments by the European Twisted Mass Collaboration in extracting parton distribution functions from the quasi-PDF approach. We concentrate on the non-perturbative renormalization prescription recently developed by us, using the RI$'$ scheme. We show results for the renormalization functions of matrix elements needed for the computation of quasi-PDFs, including the conversion to the $\MSb$ scheme, and for renormalized matrix elements. We discuss the systematic effects present in the $Z$-factors and the possible ways of addressing them in the future.
}
\begin{flushright}
DESY 17-140 
\end{flushright}\maketitle
\vspace*{-2mm}
\section{Introduction}\label{intro}

Parton distribution functions (PDFs) are one of the most important tools describing the rich internal structure of hadrons, resulting from the strong interactions of valence quarks with each other and with dynamically created sea quarks and antiquarks, via exchange of gluons.
Due to the large value of the strong coupling constant at the relevant energy scales, PDFs are non-perturbative and thus, in principle, can be studied quantitatively only in the framework of Lattice QCD.
Such an evaluation from first principles would constitute an important test of non-perturbative aspects of QCD.
Moreover, it could provide important data for experiments, in particular for the transversity PDFs (and to some extent also for polarized PDFs), which are rather poorly constrained by phenomenological fits using perturbation theory.
Even for the well-constrained experimentally unpolarized case, there are still some regions that are not kinematically accessible, such as the large Bjorken $x$ region \cite{Accardi:2016qay,Alekhin:2017kpj}.
However, since PDFs are defined on the light cone, the standard lattice Euclidean formulation allows only to compute their low Mellin moments (cf.\
\cite{Constantinou:2014tga, Constantinou:2015agp,Alexandrou:2015yqa, Alexandrou:2015xts, Syritsyn:2014saa} for recent reviews). 
Attempts to reconstruct full PDFs from these moments are bound to fail, due to both practical problems (unfavourable signal-to-noise ratio for higher moments) and theoretical ones (inevitable power divergent mixings with lower dimensional operators).

A thoroughly investigated alternative to lattice computations of the moments is the method suggested by Ji \cite{Ji:2013dva}.
This approach does not rely on the light cone definition, but instead one computes spatial matrix elements using fermion operators with a finite-length Wilson line that makes the object gauge invariant.
The infinite nucleon momentum on the light cone is, in turn, replaced by a nucleon boost in a spatial direction, conventionally the third direction ($z$).
The Fourier transform of such computed matrix elements defines \emph{quasi}-PDFs (qPDFs), which can be matched to standard light cone PDFs using perturbation theory \cite{Xiong:2013bka,Chen:2016fxx}, presently available at one-loop.
If sufficiently large nucleon momentum can be achieved, which is itself a non-trivial issue due to the signal-to-noise ratio worsening for large boosts, the matching is relatively small and the higher order corrections are small in the matching, making the extraction of PDFs from this approach robust.

There has been considerable attention devoted to Ji's approach, see e.g.\ Refs.~\cite{Lin:2014zya,Gamberg:2014zwa,Alexandrou:2015rja,Chen:2016utp,Alexandrou:2016jqi}.
A plethora of aspects has already been addressed, including effects of smearing the links in the Wilson line \cite{Lin:2014zya,Alexandrou:2015rja,Chen:2016utp,Alexandrou:2016jqi}, using momentum smearing \cite{Bali:2016lva} to improve the signal at large momenta \cite{Alexandrou:2016jqi}, nucleon boost dependence in lattice data \cite{Lin:2014zya,Alexandrou:2015rja,Chen:2016utp,Alexandrou:2016jqi} and in models \cite{Radyushkin:2017ffo,Radyushkin:2016hsy} (using the relation of PDFs and transverse momentum dependent distribution functions (TMDs)) and corrections that the finite nucleon mass induces \cite{Alexandrou:2015rja,Chen:2016utp}, as well as excited states effects (dependence on the source-sink separation in the computation of the three-point functions) \cite{Alexandrou:2015rja,Chen:2016utp}.
Theoretical aspects have also been discussed. 
In particular, the role of the Euclidean signature in the computation of quasi-PDFs has been considered \cite{Carlson:2017gpk,Briceno:2017cpo} and led to the conclusion that matrix elements obtained from the lattice are the same as ones from the LSZ formula in Minkowski space. 
Question has also been raised \cite{Rossi:2017muf} about the presence of power divergent mixings in moments of qPDFs.
However, as argued in Ref.~\cite{Ji:2017rah}, the moments of qPDFs do not exist and hence the problem of such mixings does not exist either.
An alternative generalization of light cone PDFs to finite momentum has also been proposed, leading to the so-called \emph{pseudo}-PDFs \cite{Radyushkin:2017cyf,Orginos:2017kos}.
Finally, for completeness we mention that Ji's approach can also be used to compute distribution amplitudes (DAs), as shown in Refs.~\cite{Zhang:2017bzy,Broniowski:2017wbr}.

An essential aspect of PDFs computation from qPDFs has until recently been missing, which is the issue of their renormalization.
Obviously, this is crucial from the point of view of applicability of the whole programme.
The renormalizability of qPDFs to all orders in perturbation theory was recently addressed \cite{Ji:2017oey,Ishikawa:2017faj}.
These papers used two different approaches -- the former is based on the auxiliary field formalism of Dorn~\cite{Dorn:1986dt} (which can also be used to extract renormalization functions \cite{Green:2017xeu}), while the latter considered qPDFs defined in coordinate space.
The matrix elements of qPDFs contain standard logarithmic divergences, but in addition also a power divergence is present, related to the inserted Wilson line \cite{Dotsenko:1979wR}.
At one-loop in perturbation theory, the divergence appears as a linear divergence, and was calculated in Ref.~\cite{Constantinou:2017sej} for different
fermionic and gluonic actions. 
A way to eliminate the linear divergence was also developed in Ref.~\cite{Monahan:2016bvm}, where it is made finite by using the gradient flow.
An additional aspect in qPDFs renormalization is also that qPDFs operators exhibit mixing for certain Dirac structures \cite{Constantinou:2017sej}.
The aim of our present work is to perform the renormalization procedure in a fully non-perturbative manner.
Such a procedure was introduced, for the first time, in our paper \cite{Alexandrou:2017huk} and the proposed renormalization programme was also applied in Ref.~\cite{Chen:2017mzz}.
In these proceedings, we recapitulate the results of our extensive analysis that can be found in Ref.~\cite{Alexandrou:2017huk}.

\section{Renormalization prescription}
We consider the following matrix elements with Dirac structure $\Gamma$ ($\gamma_\mu$ for unpolarized, $\gamma_\mu\gamma_5$ for helicity and $\sigma_{\mu \nu}$ 
for transversity PDFs):
\begin{equation}
\label{ME}
h_\Gamma(P_3,z)=\langle P |\bar{\psi}(0,z)\,\Gamma \,W_3(z) \psi(0,0) |P\rangle,
\end{equation}
taken between boosted nucleon states $|P\rangle$ with $P=(P_0,0,0,P_3)$, i.e.\ with the spatial momentum along the third direction.
$W_3(z)$ represents the Wilson line of length $z$ in the direction of the boost.
The qPDFs can be obtained by Fourier transforming these matrix elements.

The adopted renormalization prescription relies on the non-perturbative RI$'$~scheme \cite{Martinelli:1994ty} and proceeds along the lines of the renormalization programme developed for local operators \cite{Alexandrou:2010me}. In absence of mixing, it consists in imposing the following momentum space renormalization condition, for each Wilson line length $z$:
\begin{equation}
Z_q^{-1}\,Z_{\cal O}(z)\,\frac{1}{12} {\rm Tr} \left[{\cal V}(p,z) \left({\cal V}^{\rm Born}(p,z)\right)^{-1}\right] \Bigr|_{p^2=\bar\mu_0^2} = 1\, ,
\label{renorm}
\end{equation}
where $Z_q$ is the quark field renormalization function that satisfies:
\begin{equation}
Z_q = \frac{1}{12} {\rm Tr} \left[(S(p))^{-1}\, S^{\rm Born}(p)\right] \Bigr|_{p^2=\bar\mu_0^2}  \,.
\label{Zq_cond}
\end{equation}
In both of the above equations, the momentum $p$ is set to the RI$'$ renormalization scale $\bar\mu_0$, which we choose in such a way that its third spatial component is $P_3$.
We test two choices for the other two spatial components -- we leave them to be 0 (``parallel'' choice) or equal to $P_3$ (``diagonal'').
Previous studies of local matrix elements renormalization \cite{Alexandrou:2015sea} show that the latter choice is expected to have much smaller discretization effects, as it has a much smaller value of the ratio $\hat{P}{\equiv}\sum_\rho \bar\mu_{0_\rho}^4/\left(\sum_\rho \bar\mu_{0_\rho}^2\right)^2$.
Further, ${\cal V}(p,z)$ denotes the amputated vertex function of the operator, with ${\cal V}^{{\rm Born}}$ its 
tree-level value, e.g.\ $i\gamma_3\gamma_5 \,e^{i p z}$ for the helicity operator.
$S(p)$ is the quark propagator, which equals $S^{{\rm Born}}(p)$ at tree-level.

The above RI$'$ renormalization conditions remove all the divergences present in the matrix elements.
The linear divergence resums into a multiplicative exponential factor, $e^{-\delta m |z|/a + c |z|}$, where the operator-independent $\delta m$ is the strength of the divergence, while $c$ is an arbitrary scale \cite{Sommer:2015hea} that is fixed by the renormalization condition.
The presence of this divergence is the major difference with respect to the case of local operators.
As in the case of the latter, the standard logarithmic divergence is also removed and finite renormalization is fixed with RI$'$ equations.

Analogous conditions can be written for the case when mixing is present, e.g.\ for the vector operator with Dirac structure parallel to the direction of the Wilson line that mixes with the scalar operator.
In such case, one constructs a $2{\times}2$ mixing matrix \cite{Alexandrou:2017huk}.

Having obtained the RI$'$ scheme renormalization functions, it is desirable to convert them to the conventional renormalization scheme for phenomenology, i.e.\ the $\MSb$ scheme.
Moreover, change of the scale from $\bar\mu_0$ to an $\MSb$ scale of $\bar\mu$, typically 2 GeV, can also be peformed. 
We use the one-loop conversion factor from Ref.~\cite{Constantinou:2017sej}.
The evolution to the $\MSb$ renormalization scale $\bar\mu{=}2$ GeV is performed using the intermediate Renormalization Group Invariant scheme (RGI), see Refs.~\cite{Constantinou:2014fka,Alexandrou:2015sea} for more details. 

\begin{figure}[h]
\centering
\includegraphics[scale=0.25,angle=-90]{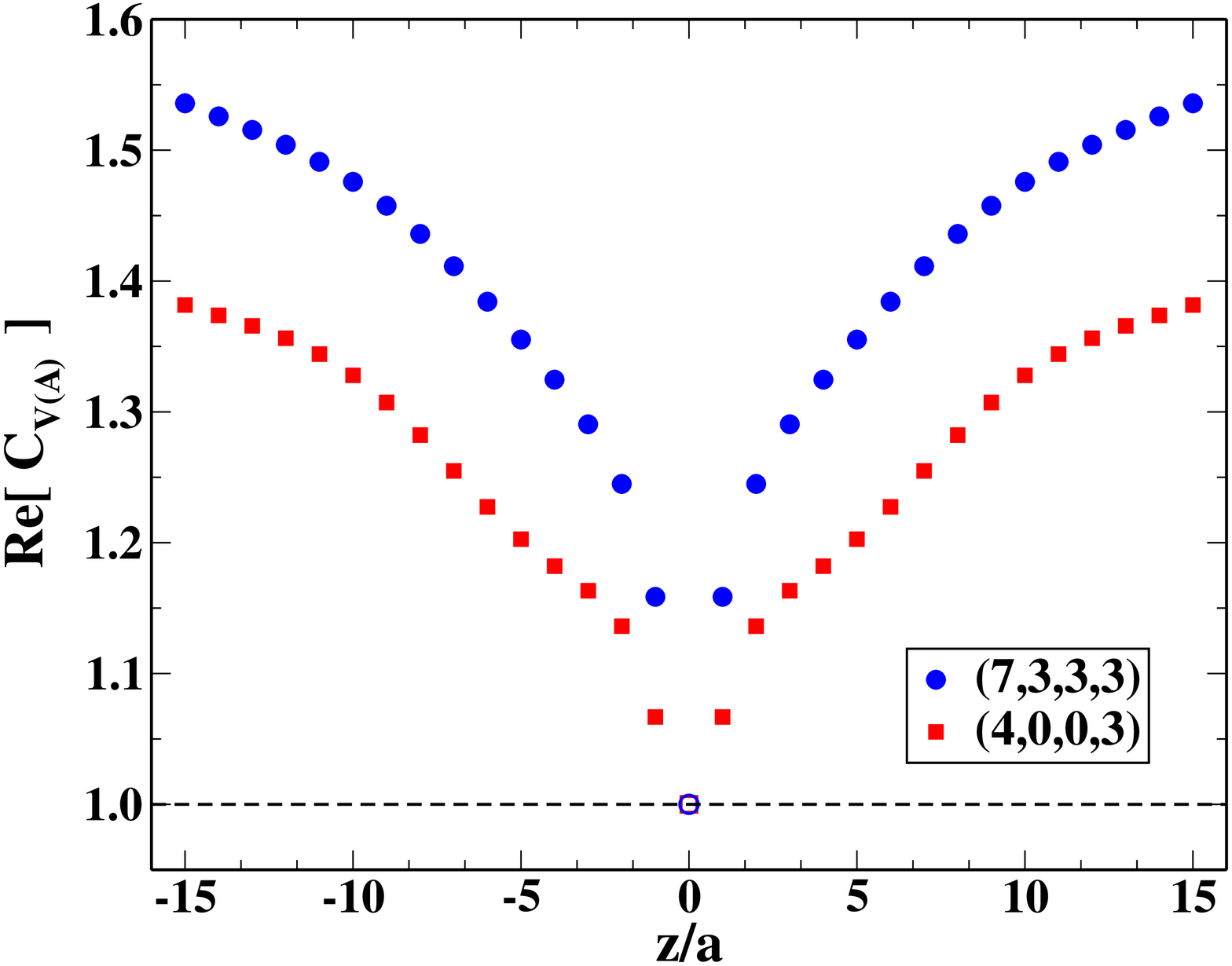}\,\,\,
\includegraphics[scale=0.25,angle=-90]{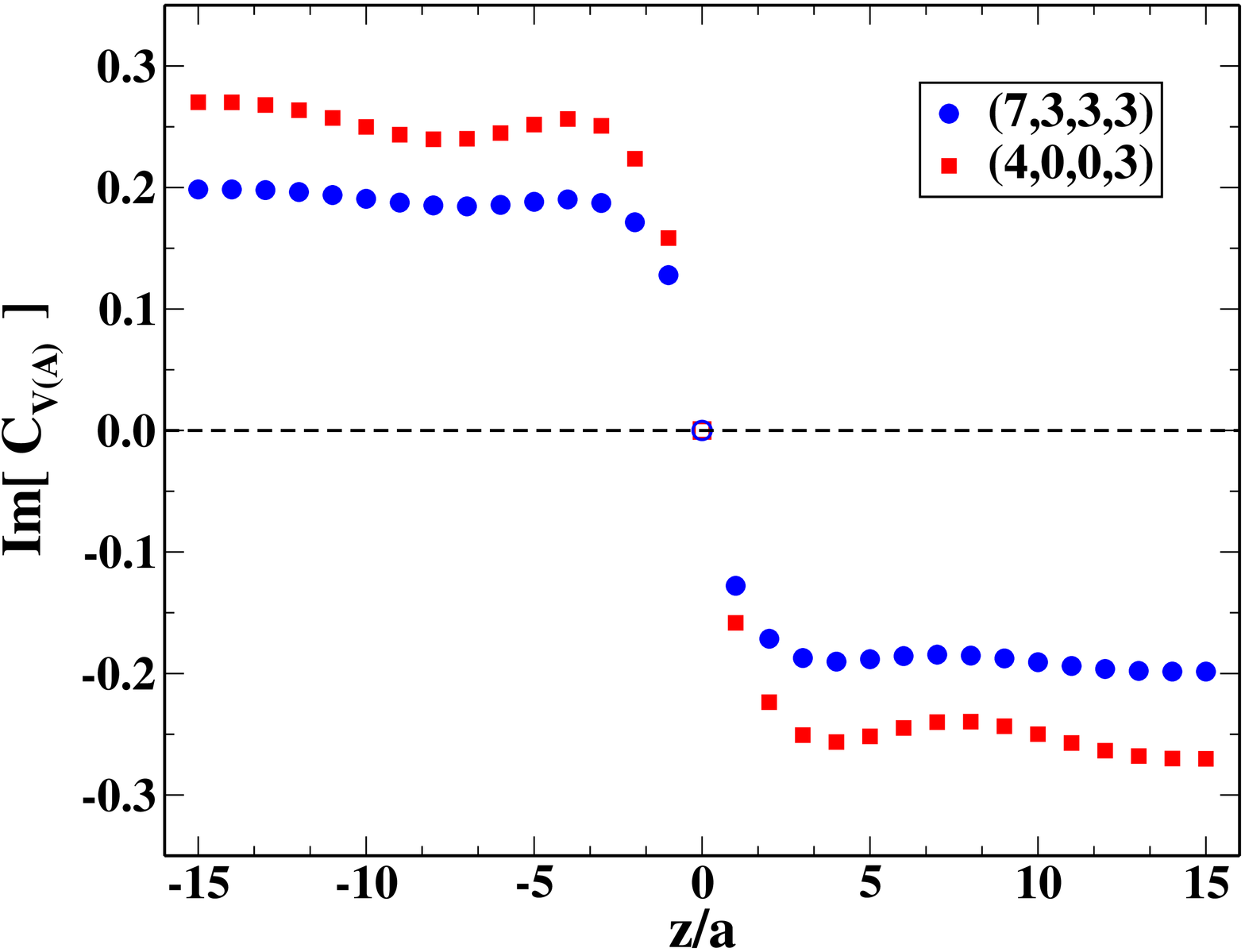}
\caption{\small{Conversion factor for the unpolarized and helicity operators from the RI$'$ scheme to the $\MSb$ scheme with $\bar\mu{=}\bar\mu_0$. 
The left plot shows the real and the right plot the imaginary part of the conversion factor vs.\ $z/a$. 
We employ two RI$'$ momenta: $a\bar\mu_0{=}\frac{2\pi}{32}\, (\frac{7}{2}{+}\frac{1}{4}, 3,3,3)$ (blue circles, labeled (7,3,3,3))  and 
$a\bar\mu_0{=}\frac{2\pi}{32}\, (\frac{4}{2}{+}\frac{1}{4}, 0,0,3)$  (red squares, (4,0,0,3))}}
\label{fig:C} 
\end{figure}

We illustrate the conversion factor for the unpolarized and helicity operators from two RI$'$ scales, $a\bar\mu_0{=}\frac{2\pi}{32}\, (\frac{7}{2}{+}\frac{1}{4}, 3,3,3)$ (``diagonal'' one) and $a\bar\mu_0{=}\frac{2\pi}{32}\, (\frac{4}{2}{+}\frac{1}{4}, 0,0,3)$ (``parallel''), to the $\MSb$ scheme at the same scale in Fig.\ \ref{fig:C}.
Each value of the conversion factor is calculated numerically (which includes the evaluation of integrals of modified Bessel functions). 
The real part is an order of magnitude larger than the imaginary part and increases with increasing $z$, whereas the imaginary part stabilizes for $|z|/a\gtrsim2$.
At $z{=}0$, the conversion equals 1, since the corresponding $Z$-factors are just ones of the local vector or axial vector currents.
For other values of $\bar\mu_0$, the behaviour of the conversion factor is qualitatively similar.

\section{Lattice setup}\label{sec:latsetup}
We use the setup of Ref.~\cite{Alexandrou:2016jqi} for calculating qPDFs matrix elements, i.e.\ we take one ensemble of $N_f{=}2{+}1{+}1$ maximally twisted mass fermions \cite{Frezzotti:2000nk,Frezzotti:2003ni,Frezzotti:2003xj,Frezzotti:2004wz} produced by the ETM Collaboration \cite{Baron:2010bv,Baron:2010th}.
The parameters of this $32^3 {\times} 64$ ensemble correspond to a lattice spacing of $a {\approx}0.082$\,fm~\cite{Carrasco:2014cwa} and a pion mass of around 375 MeV.
The nucleon boost that we use, $P_3=\frac{6\pi}{L}$, is around 1.4 GeV in physical units.
We performed 30000 measurements on 1000 configurations, using the standard Gaussian smearing of quark fields.

For the computation of the renormalization functions in the RI$'$ scheme, we employ the momentum smearing technique \cite{Gockeler:1998ye,Alexandrou:2015sea}.
This technique allows to obtain good statistical precision already with 10-20 gauge field configurations and this is the number of configurations that we have used in practice.

\section{Results}\label{sec:results}
\begin{figure}[t!]
\centering
\includegraphics[scale=0.25,angle=-90]{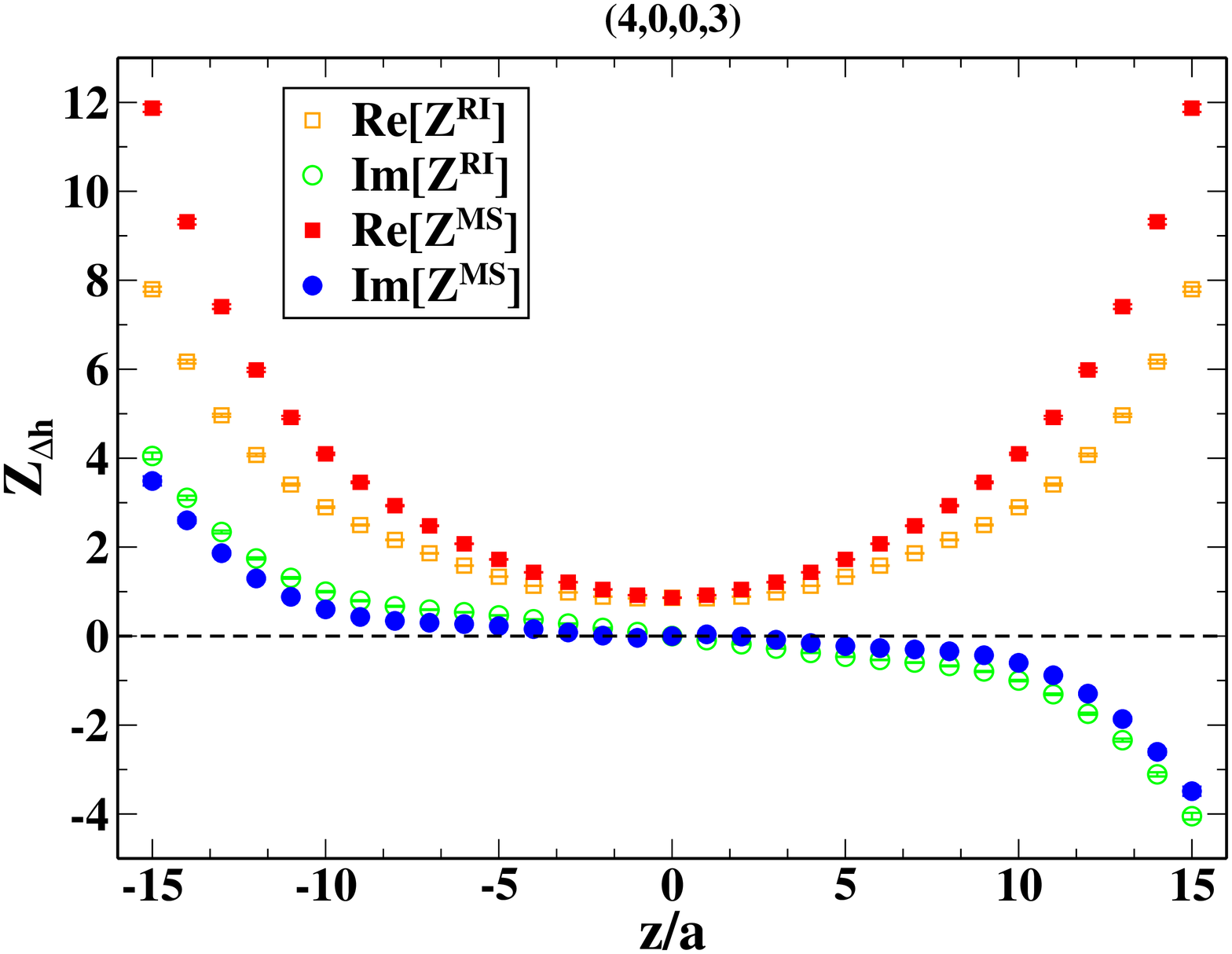}\,\,\,
\includegraphics[scale=0.25,angle=-90]{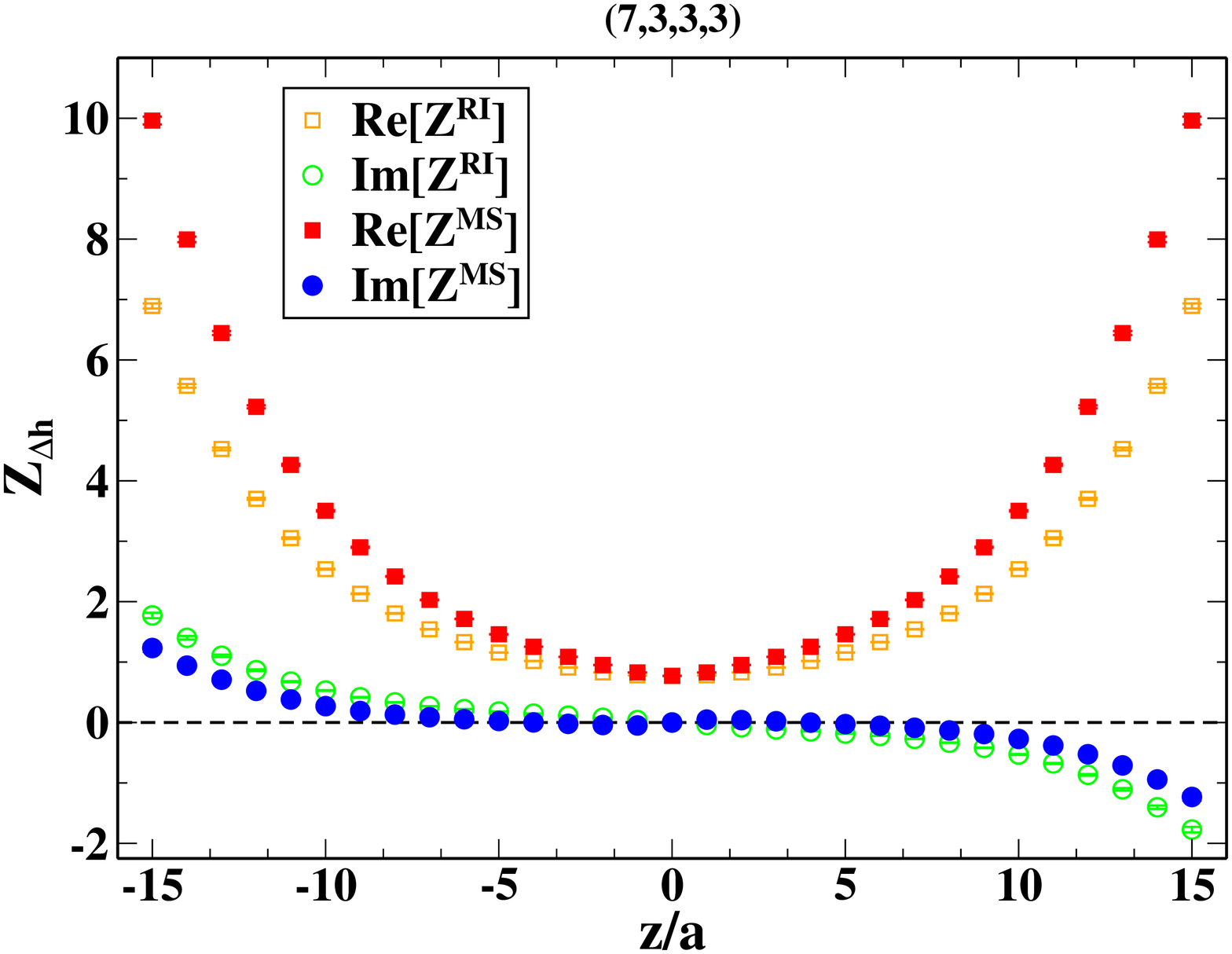}
\caption{\small{The $z$-dependence of the renormalization function for the matrix element $\Delta h(P_3,z)$ 
with $a P_3=\frac{6\pi}{L}$. We plot two cases: the ``parallel'' ($a\bar\mu_0{=}\frac{2\pi}{32}\, (\frac{4}{2}{+}\frac{1}{4}, 0,0,3)$; left) and ``diagonal'' ($a\bar\mu_0{=}\frac{2\pi}{32}\, (\frac{7}{2}{+}\frac{1}{4}, 3,3,3)$; right) choices for ${\bar{\mu}}_0$. 
The open symbols correspond to the RI$'$ scheme and the filled ones to the $\MSb$ scheme ($\bar\mu=2$ GeV).}}
\label{fig:ZA}
\end{figure}

We now show (Fig.\ \ref{fig:ZA}) the renormalization function $Z_{\Delta h}$ for the helicity case, i.e.\ the $Z$-factors that renormalize the matrix element $\Delta h(P_3,z)$. 
The left panel shows the ``parallel'' case, $a\bar\mu_0{=}\frac{2\pi}{32}\, (\frac{4}{2}{+}\frac{1}{4}, 0,0,3)$, while the right panel corresponds to the ``diagonal'' choice of the renormalization scale, $a\bar\mu_0{=}\frac{2\pi}{32}\, (\frac{7}{2}{+}\frac{1}{4}, 3,3,3)$.
In both cases, the real part of the $Z$-factors increases significantly for increasing Wilson length lines, in accordance with the presence of the power divergence.
The conversion to the $\MSb$ scheme further enhances the values.
The ``parallel'' case leads to slightly larger values of the real parts.
We can expect that these values are more contaminated by lattice artifacts in the ``parallel'' case.
This can be argued on general grounds from the higher value of $\hat{P}$, which is around 0.54, as compared to approx.\ 0.26 for ``diagonal''.
Moreover, we can compare the local value, for $z=0$, when the $Z$-factor is real and scheme- and scale-independent.
For the ``parallel'' case, we find $Z_{\Delta h}(0)=0.8620(15)$, as compared to $Z_{\Delta h}(0)=0.7727(2)$ for the ``diagonal'' choice.
Both values should be compared to the value of $Z_A{=}0.7556(5)$ found in Ref.~\cite{Alexandrou:2015sea}, where a robust calculation has been performed, including usage of several scales, subtraction of lattice artifacts computed in lattice perturbation theory and extrapolation to the limit $(ap)^2\rightarrow0$.
It is clear that the value that we find for the ``diagonal'' case is much closer to this reference value and hence, this choice of the renormalization scale $a\bar\mu_0$ is subject to smaller discretization effects.

For the imaginary parts of $Z_{\Delta h}$ in the RI$'$ scheme, we find smaller values in the ``diagonal'' case as compared to the ``parallel'' one and they are further reduced after conversion to the $\MSb$ scheme.
Again, the difference between the two choices suggests that the ``diagonal'' choice is subject to smaller lattice artifacts, since the perturbative $Z$-factor in dimensional regularization is extracted only from the poles and hence it is real to all orders in perturbation theory.
As such, the non-zero value of the imaginary part of $Z_{\Delta h}$ is an indication of both lattice artifacts and truncation effects resulting from the usage of only one-loop conversion between the schemes.

The above discussion makes it clear what needs to be done to make the extraction of $Z$-factors more robust for our case of interest.
The subtraction of lattice artifacts in lattice perturbation theory has been shown to be very efficient for the local $Z$-factors case \cite{Alexandrou:2015sea} and is expected to work in a similar fashion here.
In this way, we expect that after performing this subtraction, different scales will lead to similar values of the $Z$-factors, i.e.\ the slope of the $(ap)^2$-dependence of $Z$-factors will be significantly reduced. 
The other unwanted effect in the $\MSb$ renormalization functions comes from truncating the conversion factor to one-loop in perturbation theory.
We plan a two-loop computation to investigate this aspect.

\begin{figure}[t!]
\centering
\includegraphics[scale=0.25,angle=-90]{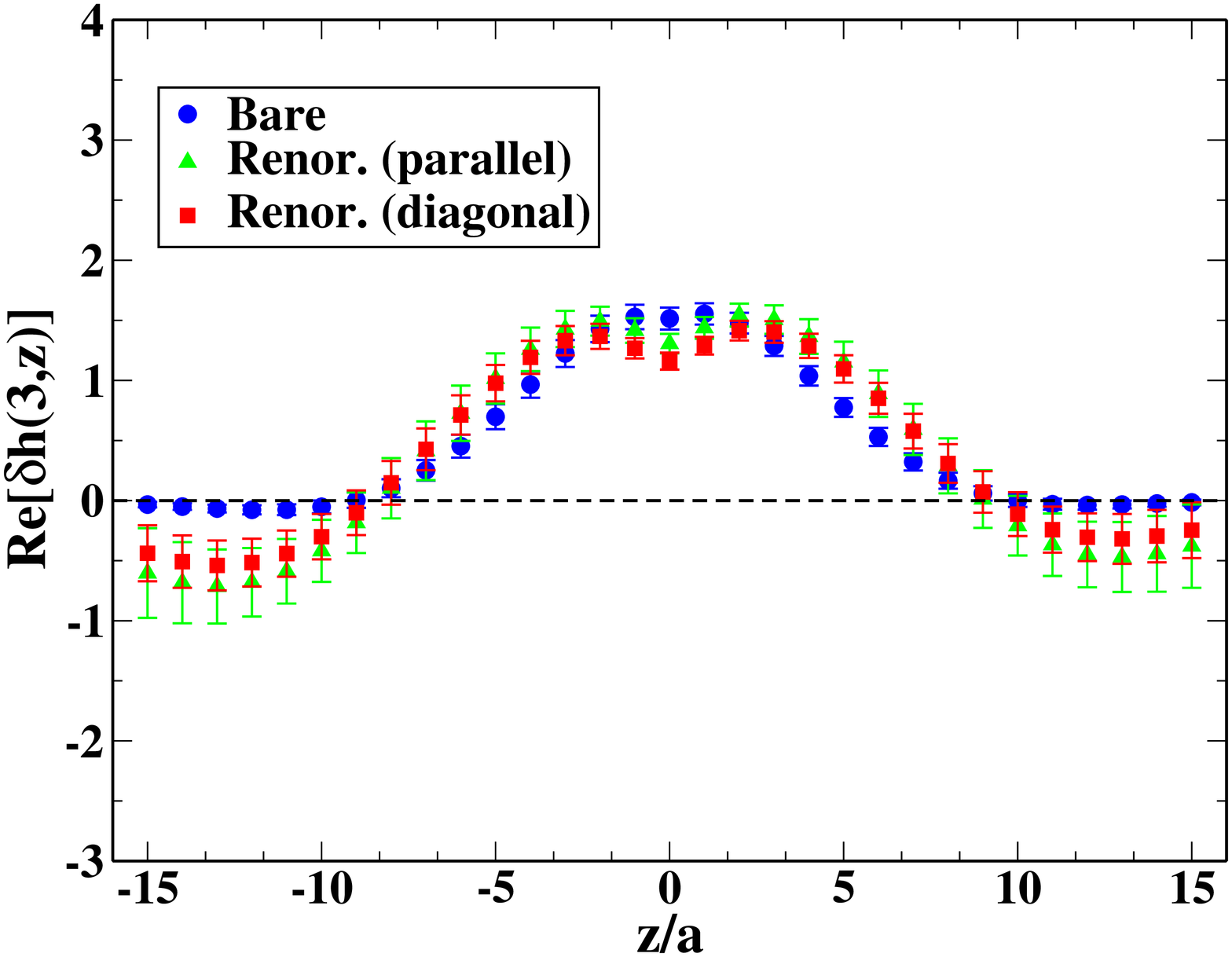}\,\,\,
\includegraphics[scale=0.25,angle=-90]{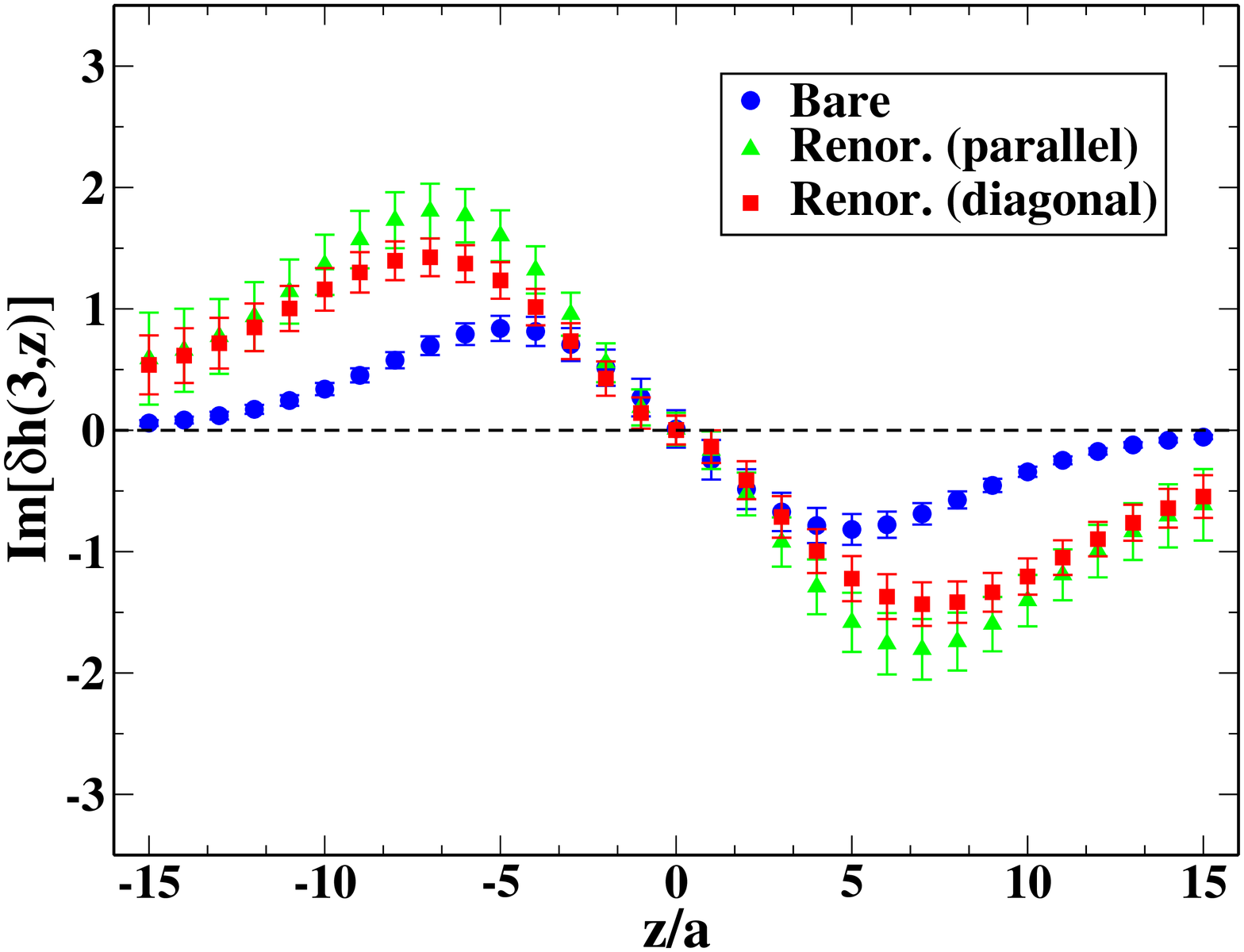}
\caption{\small{The $z$-dependence of the bare and $\MSb$ ($\bar\mu=2$ GeV) renormalized matrix elements $\Delta h(P_3,z)$ 
with $a P_3=\frac{6\pi}{L}$. The left/right panel shows the real/imaginary part.  
The blue circles correspond to the bare matrix elements, green triangles to the ``parallel'' ($a\bar\mu_0{=}\frac{2\pi}{32}\, (\frac{4}{2}{+}\frac{1}{4}, 0,0,3)$) choice of the scale and red squares to the ``diagonal'' ($a\bar\mu_0{=}\frac{2\pi}{32}\, (\frac{7}{2}{+}\frac{1}{4}, 3,3,3)$) choice.}}
\label{fig:ME}
\end{figure}

Finally, we employ the above discussed $Z$-factors to renormalize the helicity matrix elements.
In Fig.\ \ref{fig:ME} left/right, we show the real/imaginary part of bare matrix elements (blue circles), together with $\MSb$ ($\bar\mu=2$ GeV) renormalized matrix elements for the two choices of the renormalization scale, the ``parallel'' one (green triangles) and the ``diagonal'' one (red squares).
The differences between the latter are a measure of lattice artifacts present in the $Z$-factors.
Note that these differences can be treated as an upper bound of these artifacts, since they involve two very different choices of renormalization scales.
In typical extractions of the renormalization functions, one typically restricts oneself to scales with corresponding values of $\hat{P}$ below 0.35-0.40 \cite{Alexandrou:2015sea}.
In such case, lattice artifacts are much smaller and moreover, they can be reliably subtracted with a one-loop lattice perturbative calculation.
As we argued above, another source of uncertainty in the $Z$-factors and hence also in the renormalized matrix elements is the truncation of the conversion factor between the RI$'$ and $\MSb$ schemes to one-loop.
This manifests itself, for example, in the fact that the real part of renormalized matrix elements becomes negative for large Wilson line lengths.
This behaviour can occur even when the bare real part has already decayed to zero (as happens around $|z|/a=9$), but when the imaginary part is still non-zero and it is multiplied by the non-zero imaginary part of the $\MSb$ $Z$-factor.
Obviously, both of the unwanted effects would propagate to the extracted quasi-PDFs and would contaminate the extracted light-front PDFs.
Hence, addressing them is of utmost importance for the robustness of the whole research programme.

After the LATTICE 2017 conference, we have indeed addressed the systematic effects shortly described here and we refer to our paper \cite{Alexandrou:2017huk} for more details.
In short, we have investigated there several ``diagonal'' scales and estimated the magnitude of lattice artifacts as well as of truncation effects.
The explicit computation of these effects, i.e.\ the one-loop evaluation of lattice artifacts in lattice perturbation theory and calculation of the two-loop RI$'$-$\MSb$ conversion formulae is the necessary next step.

\section{Conclusions and future directions}
In this proceeding, we have summarized our prescription for the non-perturbative renormalization of matrix elements involved in the computation of quasi-PDFs.
The employed RI$'$ scheme correctly handles both kinds of divergences that are present, i.e.\ the standard logarithmic divergence as well as the Wilson line related power divergence.
We have shown our results for the RI$'$ and $\MSb$ renormalization functions for the relevant Wilson line lengths and we have discussed the systematic effects present in these $Z$-factors.
Addressing them is a necessary step to obtain robust results for the appropriate matrix elements and, thus, also for quasi-PDFs and the light-front PDFs as the final step.
In Ref.\ \cite{Alexandrou:2017huk}, we have made first quantitative assessment of such effects, i.e.\ lattice artifacts resulting from the breaking of the rotational symmetry and truncation effects in the perturbative conversion to the phenomenologically widely used $\MSb$ scheme.
However, the explicit computation of these effects remains essential.

Obviously, the systematic effects that need to be addressed for a robust extraction of light-front PDFs are not restricted to the ones present in the renormalization functions.
Apart from this, several other effects have to be taken into account and this direction is also currently pursued by us.
Arguably the most important of these effects is the influence of the pion mass.
So far, our computations \cite{Alexandrou:2015rja,Alexandrou:2016jqi,Alexandrou:2017huk} have focused on an ensemble of $N_f=2+1+1$ twisted mass fermions at a non-physical pion mass of around 375 MeV.
However, as shown e.g.\ in Ref.\ \cite{Abdel-Rehim:2015owa}, the moments of PDFs with such a pion mass are considerably different than the experimentally measured ones.
Moreover, the moments extracted from the 375 MeV ensemble using the quasi-PDF approach \cite{Alexandrou:2016jqi} (including matching, but with renormalization proxied only by HYP smearing and the usage of the local currents $Z$-factor) agree with the extraction of Ref.\ \cite{Abdel-Rehim:2015owa} within uncertainties.
This indicates that the pion mass can indeed considerably influence the shape of the quasi-distributions.
To investigate these effects, we are currently performing computations for an ensemble of $N_f=2$ twisted mass fermions with a clover term, at the physical pion mass, see Ref.\ \cite{Aurora} in these proceedings.

\vspace*{0.5cm}
\centerline{\bf\large{Acknowledgements}}
\medskip
We thank all members of ETMC for fruitful discussions. 
We also thank Rainer Sommer for discussions on the arbitrary scale in the renormalization prescription. 
KC was supported in part by the Deutsche Forschungsgemeinschaft (DFG), project nr. CI 236/1-1.
MC acknowledges financial support by the U.S. Department of Energy, Office of Science, Office of Nuclear Physics, 
within the framework of the TMD Topical Collaboration.
We acknowledge funding from the European Union's Horizon 2020 research and innovation program
under the Marie Sklodowska-Curie grant agreement No 642069.

\bibliography{references}

\end{document}